\documentstyle[aps,preprint,prl]{revtex}
\begin{document}
\draft
\preprint{}

\title{Precise Measurement of Optical Reflectivity Spectra Using
Defect-Free Surface of La$_{1-x}$Sr$_x$MnO$_3$: Evidence Against
Extremely Small Drude Weight}

\author{K.Takenaka,\cite{take} K.Iida, Y.Sawaki, and S.Sugai}
\address{Department of Physics,
Nagoya University, Nagoya 464-8602, Japan}

\author{Y.Moritomo and A.Nakamura}
\address{Center for Integrated Research in Science and Engineering
(CIRSE), Nagoya University, Nagoya 464-8603, Japan}

\date{Received 3 August 1998}

\maketitle

\begin{abstract}
Optical reflectivity spectra of La$_{1-x}$Sr$_x$MnO$_3$
(0$\leq$$x$$\leq$0.30) were measured on {\it cleavage} surfaces of
single crystals. The optical conductivity $\sigma(\omega)$ of
ferromagnetic-metal La$_{0.70}$Sr$_{0.30}$MnO$_3$ is characterized by
a Drude-like $-$ but not simple-Drude $-$ component with large
spectral weight below $\sim$1.6 eV, which yields a large effective
carrier number $N^*_{\rm eff}$ consistent with the results of Hall
coefficient and specific heat measurements. The present result
demonstrates that the previous result of \lq\lq small Drude weight"
originates from the damage of the sample surface. The
Sr-substitution effect on the electronic structure and the origin of
the unconventional $\sigma(\omega)$ are also discussed.
\end{abstract}


\narrowtext
The motivation of the renewed interest in the double-exchange
ferromagnetic-metal manganites has two $-$ fundamental and practical
$-$ aspects: study on anomalous metallic phase near Mott transition
and industrial application of the intriguing phenomena induced by a
magnetic field such as \lq\lq colossal" magnetoresistance
(CMR)\cite{cmr}. Optical reflectivity study can make an essential role
in the both stages because it enables us to not only deduce dielectric
function but also examine separately two key-elements of charge
transport $-$ carrier density (or Drude weight) and scattering time. 
The charge transport is one of the central concerns among the both
fields.

The previous reflectivity studies on La$_{1-x}$$A_x$MnO$_3$ ($A$
$=$ Sr\cite{okimoto97}, Ca\cite{kim97,jung97,jung98})
have revealed the outline of the systematic change of the electronic
structure with Sr(Ca)-substitution and the conspicuous transfer of
spectral weight over a wide energy range of 0-3 eV accompanied
by the spin polarization of conducting $e_g$ electrons. However, a
quantitative conclusion is difficult to be drawn on the far-infrared
response and on the estimation of optical (dielectric) functions
because the experiments were made using polycrystals or/and did not
cover the sufficient energy region. Especially, the origin or truth
of \lq\lq small Drude weight" $-$ pointed out by Okimoto
{\it et al.}\cite{okimoto97} but contradictory to the results of Hall
coefficient\cite{asamitsu98}, specific
heat\cite{coey95,woodfield97}, and optical absorption
studies\cite{moritomo97} $-$ is still not well-investigated.

We report the optical reflectivity spectra $R(\omega)$ of
La$_{1-x}$Sr$_x$MnO$_3$ (0$\leq$$x$$\leq$0.30) measured on the {\it
cleavage} surfaces of single crystals for a wide energy range
covering far-infrared(far-IR) region down to 5 meV. Our careful
experiment enables us to obtain the established optical functions $-$
optical conductivity $\sigma(\omega)$ and Drude weight $-$ for this
prototypical double-exchange system. Especially the Drude weight
estimated for $x$$=$0.30 is {\it large} and consistent with the
results of Hall effect and specific heat studies.

Single crystals of La$_{1-x}$Sr$_x$MnO$_3$ were grown by a floating
zone method. The details of the growth condition and characterization
were described elsewhere \cite{urushibara95}. In order to obtain
defect-free surfaces we cleaved the single-crystalline rod. The
typical size of the cleavage surface was 0.2 mm$\times$0.2 mm, which
was sufficient for optical measurements when we used the microscope
designed for the infrared-visible spectrometer.

Near-normal incident reflectivity measurements were made using a
rapid scanning Fourier-type interferometer (0.005-2.2 eV), a grating
spectrometer (1.2-6.5 eV), and a Seya-Namioka type spectrometer for
vacuum-ultraviolet synchrotron radiation (4.0-40 eV) at the Institute
for Molecular Science, Okazaki National Research Institutes. All the
data were taken at room temperature (295 K). The experimental error
of reflectivity measurement $\Delta R$ is less than 2\% over the
energy range covered here.

Figure 1 shows the reflectivity spectra measured on the cleavage
surfaces of La$_{1-x}$Sr$_x$MnO$_3$ (0$\leq$$x$$\leq$0.30) single
crystals on a logarithmic energy scale in the range 0.01-6.0 eV. The
curves are shifted upwards as Sr-composition increases. The Curie
temperature $T_{\rm C}$ of the sample is measured by dc resistivity: 
$T_{\rm C}$$=$235 K ($x$$=$0.15), 283 K (0.175), 305 K (0.20), and
362 K (0.30), yielding $T/T_{\rm C}$$=$ 1.26, 1.04, 0.967, and 0.815, 
respectively. The insulator-metal(I-M) phase boundary crosses room
temperature (295 K) between $x$$=$0.175 and 0.20 in the present case.

Undoped LaMnO$_3$ is a charge-transfer(CT) insulator in the
Zaanen-Sawatzky-Allen scheme\cite{zaanen85}. The reflectivity
spectrum exhibits several sharp peaks in the far-IR region due
to optical phonons and two broad peaks, at about 2 eV and 5 eV, 
which are assigned to the CT excitations, 
$t^3_{2g}e^1_g$$\rightarrow$$t^3_{2g}e^2_g\underline{L}$ and
$t^3_{2g}e^1_g$$\rightarrow$$t^4_{2g}e^1_g\underline{L}$
($\underline{L}$ denoting a ligand hole), respectively\cite{arima95}. 
With slight substitution ($x$$\le$0.10), the lower CT
peak disappears and a reflectivity edge suddenly appears at
$\sim$1.6 eV. As $x$ increases, the edge becomes sharpened, though
its position does not shift appreciably, and the optical phonons
are screened. For $x$$=$0.30 (ferromagnetic-metal), the optical
phonons almost fade away and the spectrum is characterized by a sharp
edge and a large spectral weight below it.

In order to make quantitative discussion on the implications of the
reflectivity data, we deduce optical conductivity $\sigma(\omega)$
(Fig. 2) from $R(\omega)$ shown in Fig. 1 via Kramers-Kronig(K-K)
transformation. Since the experiment covers the region up to 40 eV
which includes the contributions from almost all the valence
electrons in the material, the extrapolation to higher energy region
$-$ we assumed $R$$\propto$$\omega^{-4}$ $-$ is not the matter
concerning the result in the energy region of interest. For the
extrapolation to energies below 5 meV, we assumed constant $R(\omega)$
for $x$$=$0$-$0.175. For $x$$=$0.20, we make a smooth extrapolation
using Hagen-Rubens(H-R) formula with $\sigma(0)$$\sim$100
$\Omega^{-1}$cm$^{-1}$, which is roughly in accord with the dc
value. For $x$$=$0.30, on the other hand, there is some room for
varying extrapolated reflectivities (Fig. 3). The value of the dc
resistivity for $x$$=$0.30 in the present experiment is 620$\pm$50
$\mu$$\Omega$cm at 295 K, corresponding to
$\sigma (0)$$\sim$1500$-$1750 $\Omega^{-1}$cm$^{-1}$. In this range, 
a simple H-R extrapolation (dashed lines) appears not valid. However, 
this discrepancy can be removed by an extrapolation of Drude formula; 
$\epsilon(\omega)=\epsilon_{\infty}-[4\pi\sigma(0)/\omega
(\omega\tau+i)]$ ($\epsilon_{\infty}$: dielectric constant in the
high energy limit; $\tau$: scattering time)\cite{wooten}. We
assumed $\epsilon_{\infty}$$=$5 here\cite{drude}. Even in the case of
the lowest $\sigma(0)$ value (1500 $\Omega^{-1}$cm$^{-1}$), we obtain
rather smooth extrapolation by setting $\hbar/\tau$$\sim$500 cm$^{-1}$
(dashed-and-dotted line). One of the most smooth extrapolations is
the case for $\sigma(0)$$=$1700 $\Omega^{-1}$cm$^{-1}$ and
$\hbar/\tau$$=$2000 cm$^{-1}$ (solid line). Although the value of
2000 $\Omega^{-1}$cm$^{-1}$ itself has less physical meaning, the
smooth extrapolation of Drude formula suggests suppression of the
scattering rate for $x$$=$0.3 at 295 K\cite{hr}. At any case, 
variation of the extrapolation procedures was confirmed to cause
negligible difference for $\sigma(\omega)$ above 0.01 eV.

Strontium-substitution affects the optical conductivity for a wide
energy range up to $\sim$6 eV. The lower CT peak ($\sim$2.4 eV)
disappears immediately and the higher CT peak ($\sim$5.2 eV)
gradually decreases and shifts downwards as Sr-substitution
proceeds. The reduced spectral weight at the two CT peaks is
transferred to the lower energy region for the doped manganites. For
$x$$=$0.15$-$0.20, $\sigma(\omega)$ spectrum is characterized solely
by a broad peak centered at about 1 eV, which gradually develops and
the peak position shifts downwards as $x$ increases; even for
$x$$=$0.20, where the material is ferromagnetic-metallic, a
Drude-like component is not observed. For $x$$=$0.30, the spectrum
exhibits a large continuum below the edge ($\sim$1.6 eV) centered at
$\omega$$=$0. The transfer of the ${\rm spectral}$ weight is
justified by the trend that the curves of integrated spectral weight
$N^*_{\rm eff}(\omega)$ (inset of Fig.2) given as

$$N^*_{\rm eff}(\omega)={{2m_0V} \over {\pi e^2}} \int_0^\omega
\sigma(\omega ')d\omega ' \eqno(1)$$

\noindent
($m_0$: a bare electron mass; $V$: the volume per Mn-atom) merge
into a single line above 6 eV. Imperfect convergence is most likely
due to the transfer of the spectral weight in the higher-energy
region above 6 eV or/and the increasing experimental error on
$\sigma(\omega)$ with $\omega$.

The immediate vanishing of the lower CT peak indicates that the
final state of this excitation, $e_g$$\uparrow$, merges into the
valence band, strongly hybridized O$2p$$-$Mn$e_g$$\uparrow$ orbital, 
and builds up one conduction band with hole-like Fermi surface. This
interpretation is supported by (1) insensitiveness of reflectivity
edge to the doping content and by (2) positive and weakly
$x$-dependent Hall coefficient(carrier density $n$ being estimated to
be $\sim$1 hole/Mn-atom)\cite{asamitsu98}. The downward shift of
the higher CT peak is explained by the change of Madelung potential
in the Sr-substituted lattice\cite{okimoto97}. However, the decrease
of the intensity is hardly interpreted by the change of Madelung
potential alone. It is necessary to take itinerancy of the carriers
into consideration, {\it i.e.} the evolution of the spin-polarization
reduces the initial states of the second CT excitation.

One of the main purposes in the present study is the estimation of
the \lq\lq Drude weight". In a simple Drude model, $N^*_{\rm
eff}(\omega_{\rm ps})$$=$$n(m_0/m^*)$ ($n$, $\omega_{\rm ps}$, and
$m^*$ being carrier number per Mn-atom, plasma frequency, and
effective mass, respectively). $N^*_{\rm eff}(\omega_{\rm ps})$ is
so-called Drude weight. It is difficult, however, in the present
case to estimate Drude weight because a simple Drude model seems not
valid; the decay rate is too slow compared with $\propto$$\omega^{-2}$
and hence the spectral weight at the mid- and near-IR regions is too
large to be interpreted within the Drude model. Consequently, there
is some arbitrariness in the choice of $\omega_{\rm ps}$ and/or
a possibility that estimated Drude weight contains both coherent and
incoherent parts. However, the present result clearly demonstrates at
least the following fact: for $x$$=$0.30 Drude weight is much larger
than that of the previous work\cite{okimoto97}. For example, we assume
$\omega_{\rm ps}$$=$1.6 eV (reflectivity edge), then
$N^*_{\rm eff}(\omega_{\rm ps})$$=$0.346. Using $n$$\sim$1 per
Mn-atom estimated from Hall effect study\cite{asamitsu98}, 
mass-enhancement $m^*/m_0$ is estimated to be 2.9, which is
consistent with the result of specific heat
measurement\cite{coey95,woodfield97}. The present result clearly
{\it denies} the previous picture that the optical response is
characterized by such small Drude weight that we cannot attribute it to
the mass-enhancement. Although the unconventional nature of
$\sigma(\omega)$ indicates that even for $x$$=$0.30 the common
mechanism is persistently working which yields the incoherent broad
peak for the compositions near I-M transition, such mechanism is
considered to be weakened as $x$ increases.

The above mechanism is hardly produced solely by the
double-exchange interaction\cite{furukawa95}. The most plausible
candidate of the additional physics is Jahn-Teller instability. The
model incorporating both double-exchange and dynamical Jahn-Teller
couplings\cite{millis96} predicts that the carriers are localized as
polarons at $T$$>$$T_{\rm C}$ or where electron-phonon coupling
constant $\lambda$ is large, but restore gradually the metallic
behavior as $T$ or $\lambda$ decreases. In this scenario, the broad
peak observed for $x$$=$0.15$-$0.20 is interpreted due to the small
polarons. The formation of the polarons is supported also by the
optical\cite{moritomo97,kaplan96} and x-ray\cite{lanzara98}
absorption studies. It is difficult, however, to make a
quantitative comparison between the experimental results and the
theoretical predictions. This is because the nominal
concentration $x$ does not correspond to the carrier density (or
doping level) $n$ straightforward. Moreover, the Sr-substitution is
considered to increase the band-width $W$ or/and weaken the
Jahn-Teller instability. In fact, the decrease of $x$ in the actual
spectra $\sigma(\omega)$ seems to correspond to the decrease of
$\lambda$ rather than the decrease of $n$ in the calculated spectra, 
though the calculated spectra reproduce the actual spectra as a whole
(Fig. 7 of Ref.\cite{millis96}).

Another candidate is the orbital degree of freedom. One of the
theories with emphasis on it is the \lq\lq orbital liquid
theory"\cite{ishihara97}, where the isospin is introduced to describe
the orbital degree of freedom. The charge dynamics is characterized
by a crossover around $\Omega$ corresponding to the tunneling
frequency in the vibronic state of the isospin coupled with the
Jahn-Teller distortion: the orbital fluctuation behaves like a
quasi-static disorder for $\omega$$>$$\Omega$ and like an averaged
potential during a lot of tunneling processes for $\omega$$<$$\Omega$
and hence yields the incoherent and coherent responses, respectively. 
The Fermi-liquid picture incorporating the orbital degree of freedom
is also proposed\cite{shiba97} and it predicts that the orbital
dependence of transfer integral produces the interband transition
within $e_g$ orbitals. At any case, however, these theories should
predict the more detailed $x$-dependence of $\sigma(\omega)$ in order
to clarify the effect of the orbital degree of freedom on the charge
dynamics.

Finally we show that the discrepancy to the previous result
originates from the damages of the sample surface introduced by
polishing. In Fig. 4 are shown the reflectivity spectra measured on
the cleavage surface (solid line) as well as that measured on the
surface polished by lapping films with diamond powder of diameter 0.5
(dashed line) or 0.3 (dashed-and-dotted line) $\mu$m. It is found
that polishing distorts drastically $R(\omega)$ for ferromagnetic
metal La$_{0.70}$Sr$_{0.30}$MnO$_3$ [Fig. 4(b)] whereas it alters
little for undoped LaMnO$_3$ [Fig. 4(a)]. The previous data by Okimoto
{\it et al.}\cite{okimoto97} resembles closely the spectrum measured
on the surface polished by 0.3 $\mu$m-diamond film. The damage of the
surface probably localizes the carriers. However, as the wavelength
of the incident light becomes longer, light reaches the inner, not
damaged part, and hence $R(\omega)$ recovers the true spectrum, which
is consistent with that the discrepancy almost disappears below 0.03
eV. \lq\lq Small Drude weight" originates from the above restoration
process [inset of Fig. 4(b)]\cite{sdw}.

In summary, we have established the optical response of the
prototypical double-exchange system La$_{1-x}$Sr$_x$MnO$_3$ by the
precise measurement of optical reflectivity spectra using the
cleavage surfaces. For the compositions near the insulator-metal
transition, $\sigma(\omega)$ is characterized solely by an incoherent
broad peak at about 1 eV. However, the anomaly weakens as
Sr-substitution proceeds and $\sigma(\omega)$ of ferromagnetic-metal
La$_{0.70}$Sr$_{0.30}$MnO$_3$ exhibits a pronounced Drude-like, but
not simple-Drude, component, which yields the large Drude weight
consistent with the results of Hall effect and specific heat studies. 
The present result demonstrates that polish of the sample surface
damages optical spectra of the doped manganites, suggesting that
the charge dynamics of the conducting electrons in the manganites are
extremely sensitive to the static imperfections.

The authors would like to thank Professor M.~Kamada and
Dr.~M.~Hasumoto at UVSOR facility of Institute for Molecular Science, 
Okazaki National Research Institutes, for the useful advice and
skillful technical assistance to use vacuum-ultraviolet radiation. 
They are also pleased to acknowledge the help of R.~Yamamoto, 
S.~Kashima, and M.~Suzuki for the experiments. This work was partly
supported by Grant-in-Aid for Scientific Research from the Ministry
of Education, Science, and Culture of Japan and by CREST of JST.

\begin{figure}
\caption{Optical reflectivity spectra measured on the ${\rm cleavage}$
surface of La$_{1-x}$Sr$_x$MnO$_3$ (0$\leq$$x$$\leq$0.30). All data
were taken at room temperature (295 K). The curves are shifted
upwards with Sr-composition increases.}
\label{ref}
\end{figure}

\begin{figure}
\caption{Optical conductivity spectra of La$_{1-x}$Sr$_x$MnO$_3$
deduced from the reflectivity spectra measured on the cleavage
surfaces (shown in Fig. 1) via Kramers-Kronig transformation: 
$x$$=$0 (dashed-and-double-dotted line), 0.10 (long-dashed line), 
0.15 (dotted line), 0.175 (dashed-and-dotted line), 0.20 (short-dashed
line), and 0.30 (solid line). Inset: Effective carrier number per
Mn-atom $N^*_{\rm eff}(\omega)$ defined as the integration of
$\sigma(\omega)$. $T_{\rm C}$$=$235 K ($x$$=$0.15), 283 K (0.175), 
305 K (0.20), and 362 K (0.30). The insulator-metal phase boundary
crosses room temperature (295 K) between $x$$=$0.175 and 0.20.}
 \label{sig}
\end{figure}

\begin{figure}
\caption{Magnified reflectivity spectra of
La$_{0.70}$Sr$_{0.30}$MnO$_3$ in the far-infrared region (0-0.02 eV). 
Solid circle represents the actual data measured on the cleavage
surface. Extrapolated reflectivities using Drude or Hagen-Rubens
(dashed lines) formula are also shown. The variable parameters at
the Drude-extrapolation are as follows: $\sigma(0)$$=$1700 
$\Omega^{-1}$cm$^{-1}$, $\hbar/\tau$$=$2000 cm$^{-1}$ (solid line)
and $\sigma(0)$$=$1500 $\Omega^{-1}$cm$^{-1}$, $\hbar/\tau$$=$500
cm$^{-1}$ (dashed-and-dotted line). $\epsilon_{\infty}$ is
fixed at 5. The former is one of the most smooth extrapolation
here.}  \label{drude}
\end{figure}

\begin{figure}
\caption{Optical reflectivity spectra measured on the cleavage
(solid line) and polished surfaces: (a) LaMnO$_3$ and (b)
La$_{0.70}$Sr$_{0.30}$MnO$_3$. We polished the surfaces using lapping
films with diamond powder of diameter 0.5 (dashed line) or 0.3
(dashed-and-dotted line) $\mu$m. Inset: Optical conductivity spectra
deduced from the Kramers-Kronig transformation of the reflectivity
spectra shown in Fig. 4(b).}  \label{diamond}
\end{figure}


\begin{references}
  \bibitem[*]{take} Electronic address: k46291a@nucc.cc.nagoya-u.ac.jp

  \bibitem{cmr} For example, R. von Helmolt {\it et al}., Phys. Rev.
  Lett. {\bf 71}, 2331 (1993); Y. Moritomo {\it et al}., Nature
  {\bf 380}, 141 (1996).

  \bibitem{okimoto97} Y. Okimoto {\it et al}., Phys. Rev. B
  {\bf 55}, 4206 (1997).

  \bibitem{kim97} K. H. Kim {\it et al}., Phys. Rev. B {\bf 55},
  4023 (1997).

  \bibitem{jung97} J. H. Jung {\it et al}., Phys. Rev. B {\bf 55},
  15489 (1997).

  \bibitem{jung98} J. H. Jung {\it et al}., Phys. Rev. B {\bf 57},
  R11043 (1998).

  \bibitem{asamitsu98} A. Asamitsu and Y. Tokura, Phys. Rev. B
  {\bf 58}, 47 (1998).

  \bibitem{coey95} J. M. D. Coey {\it et al}., Phys. Rev. Lett. 
  {\bf 75}, 3910 (1995).

  \bibitem{woodfield97} B. F. Woodfield, M. L. Wilson, and
  J. M. Byers, Phys. Rev. Lett. {\bf 78}, 3201 (1997).

  \bibitem{moritomo97} Y. Moritomo {\it et al}., Phys. Rev. B
  {\bf 56}, 5088 (1997); A. Machida, Y. Moritomo, and A. Nakamura,
  Phys. Rev. B, in press.

  \bibitem{urushibara95} A. Urushibara {\it et al}., Phys. Rev. B
  {\bf 51}, 14103 (1995).

  \bibitem{zaanen85} J. Zaanen, G. A. Sawatzky, and J. W. Allen,
  Phys. Rev. Lett. {\bf 55}, 418 (1985).

  \bibitem{arima95} T. Arima and Y. Tokura, J. Phys. Soc. Jpn.
  {\bf 64}, 2488 (1995).

  \bibitem{wooten} F. Wooten, {\it Optical Properties of Solids}, 
  (Academic Press, Inc., New York, 1972).

  \bibitem{drude} For the various 3$d$-transitional metal oxides,
  $\epsilon_{\infty}$ is estimated to be 2$-$10; See Ref.\cite{jung98},
  \cite{arima95}, and T. Katsufuji, M. Kasai, and Y. Tokura, Phys.
  Rev. Lett. {\bf 76}, 126 (1996). We confirmed that the variation of
  $\epsilon_{\infty}$ from 1 to 10 yields negligible small difference
  of reflectivity below 0.01eV.

  \bibitem{hr} A Hagen-Rubens formula corresponds to a Drude formula
  in the limit $\hbar/\tau$$\rightarrow$$\infty$ and hence it is
  applicable only in the region $\omega$$\ll$$\hbar/\tau$.

  \bibitem{furukawa95} N. Furukawa, J. Phys. Soc. Jpn. {\bf 64},
  3164 (1995).

  \bibitem{millis96} A. J. Millis, R. Mueller, and
  Boris I. Shraiman, Phys. Rev. B {\bf 54}, 5405 (1996).

  \bibitem{kaplan96} S. G. Kaplan {\it et al}., Phys. Rev. Lett.
  {\bf 77}, 2081 (1996).

  \bibitem{lanzara98} A. Lanzara {\it et al}., Phys. Rev. Lett.
  {\bf 81}, 878 (1998).

  \bibitem{ishihara97} S. Ishihara, M. Yamanaka, and N. Nagaosa,
  Phys. Rev. B {\bf 56}, 686 (1997).

  \bibitem{shiba97} H. Shiba, R. Shiina, and A. Takahashi,
  J. Phys. Soc. Jpn. {\bf 66}, 941 (1997).

  \bibitem{sdw} For the bilayer manganite a Drude component was not
  observed in the optical spectra measured even on the cleavage
  surface [T. Ishikawa {\it et al}., Phys. Rev. B {\bf 57}, R8079
  (1998)]. However, this is probably due to the localization of the
  carriers, suggested by the logarithmic up-turn in dc resistivity
  at low temperatures.

\end{references}
\end{document}